# Inheritance of Solar Short- and Long-Lived Radionuclides from Molecular Clouds and the Unexceptional Nature of the Solar System

Edward D. Young[1]

[1]*Department of Earth and Space Sciences, University of California Los Angeles, 595 Charles E. Young Drive East, Los Angeles, CA 90095 (eyoung@ess.ucla.edu)*

Abstract

Apparent excesses in early-solar $^{26}$Al, $^{36}$Cl, $^{41}$Ca, and $^{60}$Fe disappear if one accounts for ejecta from massive-star winds concentrated into dense phases of the ISM in star-forming regions. The removal of apparent excesses is evident when wind yields from Wolf-Rayet stars are included in the plot of radionuclide abundances vs. mean life. The resulting trend indicates that the solar radionuclides were inherited from parental molecular clouds with a characteristic residence time of $10^8$ years. This residence time is of the same order as the present-day timescale for conversion of molecular cloud material into stars. The concentrations of these extinct isotopes in the early solar system need not signify injection from unusual proximal stellar sources, but instead are well explained by normal concentrations in average star-forming clouds. The results imply that the efficiency of capture is greater for stellar winds than for supernova ejecta proximal to star-forming regions.

## 1. INTRODUCTION

Correlations between radioactive decay mean lives ($\tau_R$, related to half life by $\tau_R = t_{1/2}/\ln(2)$) and radionuclide abundances (e.g., Wasserburg et al. 1996) provide a test of various scenarios for the provenance of solar-system rock-forming elements. In the simplest case of continuous production and radioactive decay, radionuclides with the shortest $\tau_R$ (shortest half lives) should be least abundant when their concentrations are normalized to their stable counterparts. Apparent excesses in $^{26}$Al/$^{27}$Al, $^{41}$Ca/$^{40}$Ca, and $^{60}$Fe/$^{56}$Fe relative to this expectation in particular have been taken as evidence for injection of these short-lived nuclides into the early solar system by a variety of sources, including supernovae proximal to the site of solar system formation (e.g. Ouellette et al. 2007; Ouellette et al. 2010; Gritschneder et al. 2012). Recent models have underscored the importance of enrichment of short-lived radionuclides in star-forming regions by collapse supernovae (SNe) ejecta (Vasileiadis et al. 2013) and/or by winds from rapidly rotating Wolf-Rayet (WR) or main-sequence WR progenitors (Gaidos et al. 2009; Gounelle and Meynet 2012).

The atomic $^{60}$Fe/$^{26}$Al (based on measurements of their decay products in meteorites) at the time of formation of the first solids in the solar system was ~ 0.002 while the steady-state Galactic value deduced from γ-ray decay activities of these isotopes is ~ 0.6 (Diehl et al. 2006; Diehl et al. 2010; Tang and Dauphas 2012; Wang et al. 2012). Supernova debris injection (Vasileiadis et al. 2013) does not offer an obvious explanation for the low $^{60}$Fe/$^{26}$Al. Specific scenarios involving multiple stages of enrichment, first by supernova ejecta in previous generations of star formation followed by subsequent enrichment by winds, can account for the depressed $^{60}$Fe/$^{26}$Al (Gounelle and Meynet 2012). In most cases, if not all, explanations for the relative abundances of the short-lived radionuclides in the early solar system point to extraordinary sequences of events involving supernovae or combinations of supernovae and WR stars proximal to the site of solar system formation in both space and time. Estimates for the probabilities of these occurrences are often in the single digit percentiles or less (e.g., Gaidos et al., 2009).

Recently, Jura et al. (2013) showed that the initial $^{26}$Al/$^{27}$Al ratio for the solar system of ~ 5x10$^{-5}$ is apparently similar to star-forming regions today in the Galaxy. Their conclusion is based mainly on two observations. Firstly, Jura et al. pointed out that if $^{26}$Al is produced locally by winds from massive stars in star-forming regions, the proper calculation for estimating the concentration of $^{26}$Al in these regions is to divide the Galactic mass of $^{26}$Al (between 1.5 and 2.2 $M_\odot$, Diehl et al, 2006; 2010) by the mass of H$_2$ in the Galaxy (8.4x10$^8$ $M_\odot$, Draine 2011) rather than the mass of total H (4.95x10$^9$ $M_\odot$) as is commonly done. This is because the former traces molecular clouds, the sites most likely to be enriched by winds. Converting the mass ratio to atomic abundances, and combining with solar abundances of Al (atomic Al/H = 3.5x10$^{-6}$, Lodders 2003), Jura et al. obtain a $^{26}$Al/$^{27}$Al ratio of 2 to 3x10$^{-5}$ for regions susceptible to star formation today. This value is consistent with the solar value when one corrects for the increase in total Al in the Galaxy over the last 4.6 Gyr (e.g., Huss et al. (2009) derive a factor for GCE $^{27}$Al growth of 1.62 over this interval). Secondly, Jura et al. noted that white dwarf stars polluted by impacting asteroids show elemental abundances implying rock-metal differentiation; the impactors in these extra-solar planetary systems were once molten. Since these rocks were essentially chondritic in bulk composition and small in size (see Jura et al. 2013 and references therein), making $^{26}$Al decay the only viable heat source, the minimum $^{26}$Al/$^{27}$Al required to produce melting can be calculated and is ≥ 3x10$^{-5}$ (Jura et al. 2013). It appears that melting of small rocky bodies by the decay of $^{26}$Al is common outside of the solar system. The conclusion is that star-forming regions throughout the Galaxy have complements of $^{26}$Al similar to that present in the young solar system.

The proposed concentration of $^{26}$Al in star-forming regions offers a natural explanation for the disparity in $^{60}$Fe/$^{26}$Al between the solar system and the average Galactic value. Where $^{26}$Al is produced mainly by WR and pre-WR main-sequence winds, the product $^{26}$Al can seed the parental molecular clouds. WR stars are sufficiently massive (e.g., >>25 $M_\odot$, lifetimes of several Myr) that they deposit their ejecta prior to escape from the vicinity of giant molecular clouds where stars are actively forming. By contrast, $^{60}$Fe is produced mainly from SNe from a variety of stellar masses with many of the progenitor stars living longer than the lifetimes of parental giant molecular clouds (Murray et al. 2010). Non-detection of γ-decay emission from $^{60}$Fe and the presence of resolvable γ emission from $^{26}$Al in the Cygnus star-forming region (SFR) where WR stars have been observed suggests the decoupling of these nuclides in at least one SFR (Martin et al. 2009), although the decoupling is uncertain due to the relatively high detection limit for $^{60}$Fe γ-decay there.

In this paper the consequences of the apparent heterogeneous distribution of $^{26}$Al in the interstellar medium (ISM) are explored for other radionuclides. The production of $^{60}$Fe, $^{41}$Ca, and $^{36}$Cl in particular, relative to $^{26}$Al, by stellar winds, and the latest measurements of the daughter products of these nuclides in meteorites, are used to update the plot of relative abundances of the radionuclides versus mean life by decay. The result indicates that the solar abundances of $^{26}$Al, $^{36}$Cl, $^{41}$Ca, $^{53}$Mn, $^{60}$Fe, $^{107}$Pd, $^{129}$I, $^{182}$Hf, $^{244}$Pu, $^{246}$Sm, $^{235}$U and $^{238}$U are *all* consistent with values expected from the average rate and efficiency of star formation in the Milky Way. There appears to be no compelling reason to invoke special circumstances to explain the concentrations of these nuclides. They evidently entered our solar system in concentrations consistent with those expected for molecular clouds in general.

2. SOLAR ABUNDANCES VS. MEAN LIVES

Apparent excesses of $^{26}$Al, $^{36}$Cl, $^{41}$Ca, and $^{60}$Fe in the solar system have been identified by plotting the ratios of the atomic abundances of the radionuclides to their stable counterparts, $N_R/N_S$, versus mean lifetimes, $\tau_R$, and comparing the resulting trends with expectations for the ISM (Wasserburg et al. 1996; Lodders and Cameron 2004; Jacobsen 2005; Huss et al. 2009). In such plots, the isotope ratios must be normalized to the production ratios for the nuclides, $P_R/P_S$, because the ratios of nuclides depend on the production ratios as well as on the rates of decay. For example, consider the differential equation for the abundance of a radionuclide in the ISM as a function of time (Jacobsen 2005):

$$\frac{dN_R}{dt} = \psi(t)P_R - \lambda_R N_R \tag{1}$$

where $\psi(t)$ is the time-dependent rate of nuclide injection into the ISM from stars, $P_R$ is the total production for the radionuclide ($\psi(t)P_R$ = atoms/yr), and $\lambda_R$ is the decay constant. The analogous expression for $dN_S$ is $dN_S = \psi(t)P_S \, dT$. Jacobsen (2005) showed that the solution for the ratio $N_R/N_S$ for extinct nuclides can be written as

$$\log\left(\frac{N_R}{N_S}\right) - \log\left(\frac{P_R}{P_S}\right) \simeq \log \tau_R - \log T^* \tag{2}$$

where $T^*$ is the age of the Galaxy weighted by an evolving production rate such that $T^* = T\langle\psi\rangle/\psi(T)$, $\langle\psi\rangle$ is the average rate of total production and $\psi(T)$ is the rate at Galactic age $T$. Equation (2) shows that on a plot of $\log[(N_R/N_S)/(P_R/P_S)]$ vs. $\log(\tau_R)$ one expects the short-lived nuclides to align on a line with slope of unity and an intercept controlled by the age of the Galaxy and the history of star formation.

Equation (2) is plotted in Figure 1 for the age of the Galaxy at the time the solar system formed assuming that the rate of nuclide production has been constant ($T^* = T =$ 12 Gyr – 4.6 Gyr = 7.4 Gyr). The plotting positions for the various short-lived and longer-lived radionuclides for which we have data from meteoritical materials are shown for comparison in Figure 1. Isotope ratios, production ratios, and mean life values used in Figure 1 are listed in Table 1. Here, as is customary, we use $^{232}$Th ($\tau_R = 2\times10^{10}$ yrs) as the pseudo-stable partner for the actinides. Production ratios are taken from Huss et al. (2009). These production ratios are derived from supernova-dominated yields integrated over a Salpeter stellar initial mass function (IMF) that specifies the numbers of stars $N_*$ of mass $M_*$ (i.e., $dN_*/dM_* = \beta M_*^{-2.35}$ and $\beta$ is a scaling parameter). They are obtained from equation 14 in Huss et al. combined with a primary yield of 0.016 (p. 4928, Huss et al. 2009), their $k$ value of unity, and a Galactic age of 7.4 Gyr (time at birth of the solar system). They are generally within a factor of 2 to 3 of earlier supernova-based production ratios (Jacobsen 2005 and references therein), suggesting uncertainties in $(N_R/N_S)/(P_R/P_S)$ of at least a factor of 2. Short-lived Be isotopes are not included in this analysis for the present as they are generally regarded as being irradiation products. $^{36}$Cl may also have an irradiation component and it is noteworthy that the initial solar $^{36}$Cl/$^{35}$Cl is uncertain to a factor of 2x or more (Lin et al. 2005).

With the exception of the longest-lived nuclides, represented here by $^{238}$U and $^{235}$U, solar-system radionuclides do not align with relative abundances predicted by Equation (2). Equation (2) is inadequate because it fails to incorporate the residence times of material in the dense phases of the ISM where supernova production does not occur. The rock-forming elements are mainly in dust, and dust in clouds is protected from destruction by supernova shock waves and sputtering that occurs in the diffuse interstellar medium. A finite residence time in clouds suppress $N_R/N_S$ relative to Equation (2).

Several authors have calculated the multi-phase ISM equivalents of Equation (2) (Jacobsen 2005; Huss et al. 2009). We will make use of the model by Jacobsen (2005) for a two-phase ISM because the results are given as analytical solutions that can be easily manipulated for illustration. The results from numerical models (e.g., Huss et al. 2009) are broadly comparable, however.

Following Jacobsen (2005) one can write four differential equations describing a box model for the ISM composed of dense molecular clouds (MC) and surrounding diffuse inter-cloud space (IC). The circumstance is depicted in Figure (2). The equations are

$$\frac{dN_{R,IC}}{dt} = \frac{N_{R,MC}}{\tau_{MC}} - \frac{N_{R,IC}}{\tau_{IC}} + \psi P_R - \lambda_R N_{R,IC} \quad (3)$$

$$\frac{dN_{S,IC}}{dt} = \frac{N_{S,MC}}{\tau_{MC}} - \frac{N_{S,IC}}{\tau_{IC}} + \psi P_S \quad (4)$$

$$\frac{dN_{R,MC}}{dt} = \frac{N_{R,IC}}{\tau_{IC}} - \frac{N_{R,MC}}{\tau_{MC}} - \lambda_R N_{R,MC} \quad (5)$$

$$\frac{dN_{S,MC}}{dt} = \frac{N_{S,IC}}{\tau_{IC}} - \frac{N_{R,MC}}{\tau_{MC}} \quad (6)$$

where $\tau_j$ is the residence time of material in reservoir $j$ and $N_{i,j}/\tau_j$ is the flux of nuclide $i$ from reservoir $j$. Equations (3) and (4) describe the time-dependence of the atomic abundances of radionuclide R and stable partner S in the diffuse ISM and Equations (5) and (6) are the same for the molecular cloud phase. It is important to recognize that the residence time in the cloud, $\tau_{MC}$, is a manifestation of the rate of transfer from molecular clouds to inter-cloud space. A large value for $\tau_{MC}$ signifies a slow bleeding of material from molecular clouds. At steady state, Jacobsen (2005) showed that Equations (3) through (6) can be combined to give

$$\log\left(\frac{N_{R,MC}}{N_{S,MC}}\right) - \log\left(\frac{P_R}{P_S}\right) = 2\log\tau_R - \log\left[(1-x_{MC})\tau_{MC} + \tau_R\right] - \log T^* \quad (7)$$

where $x_{MC}$ is the mass fraction of the ISM that exists as molecular clouds (~0.17 today, Draine 2011). For nuclides with mean lifetimes significantly less than molecular cloud residence times, the relationship between the left-hand-side of Equation (7) and $\log(\tau_R)$ is a line with a slope of 2, meaning that $N_R/N_S$ normalized to $P_R/P_S$ varies as $\tau_R^2$. For a given molecular cloud residence time and age of the Galaxy, Equation (7) predicts the abundances of the radionuclides relative to their stable partners and the associated production ratios.

Curves from Equation (7) representing the time of solar system formation and various values of $\tau_{MC}$ are compared with the meteorite data in Figure 1. A similar plot appears in Jacobsen (2005), in Huss et al. (2009) and in Lodders and Cameron (2004). The significant differences between Figure 1 and those earlier plots arise because of revisions to the initial solar $^{60}Fe/^{56}Fe$ (Tang and Dauphas 2012) and the decay mean lives of $^{60}Fe$ (Rugel et al. 2009) and $^{146}Sm$ (Kinoshita et al. 2012).

A striking feature of Figure 1 is that the radionuclides, both extant and extinct, can be divided into two groups, those with $(N_S/N_R)/(P_R/P_S)$ accounted for by $\tau_{MC}$ on the order of $10^8$ years ($^{238}U$, $^{235}U$, $^{146}Sm$, $^{244}Pu$, $^{182}Hf$, $^{129}I$, $^{107}Pd$, $^{53}Mn$), and those with abundances well above the predicted values for $\tau_{MC} \sim 10^8$ yrs ($^{26}Al$, $^{36}Cl$, $^{41}Ca$, and perhaps $^{60}Fe$). Lodders and Cameron (2004) referred to the trend defined by the former group as the "mean life squared" relationship in their analysis of extinct radionuclides in the solar system, and Jacobsen (2005) shows that it is well explained by Equation (7). The existence of the group of short-lived radionuclides defined by apparent anomalous enrichments in $^{26}Al$, $^{36}Cl$, $^{41}Ca$ and ~ $^{60}Fe$ in Figure 1 is often taken as evidence for special circumstances surrounding solar system formation (e.g., proximal supernovae). $^{10}Be$ would be in the latter group but is excluded for the time being for the reasons cited above.

## 3. THE EFFECTS OF WINDS

The production ratios for the data plotted in Figure 1 are based on integrations of supernova yields over the masses of stars comprising the initial mass function (IMF). Stellar winds are included in some SNe yield calculations. Woosley and Heger (2007) show that winds and the subsequent explosion of a 60 $M_\odot$ progenitor star produce roughly equal masses of $^{26}$Al, and that rapid rotation (300 km/s) doubles the yield of $^{26}$Al. Here we consider the effects of additions of nuclides by winds from rapidly rotating massive stars with $M_* > 25 M_\odot$ beginning on the main sequence (MS) and continuing into the Wolf-Rayet phase of evolution (Arnould et al. 1997; Arnould et al. 2006). Because massive O stars may not always produce vigorous SNe but instead experience fallback directly to form black holes, we consider winds independent from supernova yields (Fryer et al. 2007) (we return to this aspect in §4.3). WR stars eject mass at rates of ~ $10^{-5}$ $M_\odot$ yr$^{-1}$ for timescales of order $10^5$ years (Arnould et al. 1997). The more massive stars (> 25 $M_\odot$) go through more protracted episodes of mass loss lasting several Myr, extending the potential period of enrichment of O star environs.

The nuclides with the shortest mean lives that appear to be in excess in Figure 1 share the characteristic of being produced nearer to the surface of the star by H and He burning, resulting in expulsion in WR winds. For example, $^{26}$Al is produced by H burning and $^{36}$Cl and $^{41}$Ca mostly by core and shell He burning (Meyer 2005). $^{53}$Mn, on the other hand, fits well with the longer-lived nuclides in Figure 1 and is produced during Si burning (Clayton 2003). Rapid rotation of massive stars leads to ejection of H-burning products while still on the MS prior to the WN (nitrogen-rich hydrogen burning), WC (carbon-rich helium burning), and WO (oxygen-rich carbon burning) phases of WR evolution. Among these products is $^{26}$Al. Consequently, $^{26}$Al can be produced over several Myrs of O star evolution (Gounelle and Meynet 2012) leading up to the WR phase. $^{26}$Al ejection continues into the WN phase but ceases in the WC-WO phase as a result of destruction by He burning (Arnould et al. 2006). Other short-lived radionuclides of interest, including $^{36}$Cl, $^{41}$Ca, $^{60}$Fe, and $^{107}$Pd, are ejected in the WC phase. A useful measure of the relative masses of short-lived radionuclides produced by WR winds is obtained by using the $60 M_\odot$ progenitor calculations for solar metallicity presented by Arnould et al. (2006) and Gounelle and Meynet (2012) as an example. In those calculations the cumulative masses of nuclides ejected by winds are: $^{26}$Al ~ 7x10$^{-5}$ $M_\odot$, $^{36}$Cl ~ 7x10$^{-7}$ $M_\odot$, $^{41}$Ca ~ 8x10$^{-7}$ $M_\odot$, $^{60}$Fe ~ 3x10$^{-9}$ $M_\odot$, and $^{107}$Pd ~ 9x10$^{-10}$ $M_\odot$ (note that radioactive decay is included in these calculations). Winds will make significant contributions to the inventories of $^{26}$Al, $^{36}$Cl, and $^{41}$Ca in star-forming regions with lesser contributions to the budgets of $^{60}$Fe and $^{107}$Pd.

From Figure 1 and the preceding discussion it is evident that those nuclides that deviate from the trend reflecting a slow leak from molecular clouds characterized by $\tau_{MC}$ ~10$^8$ years are the same nuclides produced in significant abundance by winds from WR stars, with the important exception of $^{107}$Pd. A reasonable conclusion is that these nuclides deviate from the trend because supernova production alone is an underestimate of their total production in star-forming regions, yielding $(N_R/N_S)/(P_R/P_S)$ values that are too high. In order to test this hypothesis, one can calculate the missing $^{26}$Al production due to winds required to bring $^{26}$Al/$^{27}$Al into line with the majority of radionuclides, and then scale the wind contributions of $^{36}$Cl, $^{41}$Ca, $^{60}$Fe, and $^{107}$Pd to that of $^{26}$Al using the calculations by Arnould and others. If $(N_R/N_S)/(P_R/P_S)$ for all nuclides align with the 10$^8$ year trend, then wind production is a self-consistent explanation for the dispersion in the data in Figure 2.

For this analysis, we rewrite the production ratio $P_R/P_S$ for each short-lived nuclide as

$$\frac{P_R}{P_S} = \frac{\Lambda_{SNe} P_R^{SNe}}{\Lambda_{SNe} P_S} + \frac{\Lambda_W P_R^W}{\Lambda_{SNe} P_S} \qquad (8)$$

where $P_R^{SNe}$ is the radioisotope production term obtained from supernova yields (Table 1) and $P_R^W$ is production term due to pre-WR and WR winds. The terms $\Lambda_{SNe}$ and $\Lambda_W$ characterize the efficiency of production and capture of supernova and wind ejecta, respectively. For example, if winds dominate over supernovae in a given astrophysical setting, $\Lambda_W / \Lambda_{SNe}$ will be $\gg 1$. If there is a finite production due to supernovae but the products are not accessible in a star-forming region, for whatever reason, $\Lambda_{SNe} = 0$.

With the explicit separation of production by supernovae and winds, Equations (3) and (4) are rewritten:

$$\frac{dN_{R,IC}}{dt} = \frac{N_{R,MC}}{\tau_{MC}} - \frac{N_{R,IC}}{\tau_{IC}} + \psi(\Lambda_{SNe} P_R^{SNe} + \Lambda_W P_R^W) - \lambda_R N_{R,IC} \qquad (9)$$

$$\frac{dN_{S,IC}}{dt} = \frac{N_{S,MC}}{\tau_{MC}} - \frac{N_{S,IC}}{\tau_{IC}} + \psi \Lambda_{SNe} P_S^{SNe}. \qquad (10)$$

It is straightforward to verify that substitution of $\Lambda_{SNe} P_R^{SNe} + \Lambda_W P_R^W$ for $P_R$ and $\Lambda_{SNe} P_S^{SNe}$ for $P_S$ does not alter the derivation given by Jacobsen (2005) since the values for production are altered but not their role in the mass balance equations. Therefore, Equation (7) becomes

$$\log\left(\frac{N_{R,MC}}{N_{S,MC}}\right) - \log\left(\frac{P_R^{SNe} + (\Lambda_W / \Lambda_{SNe}) P_R^W}{P_S}\right) = 2\log \tau_R - \log[(1 - x_{MC})\tau_{MC} + \tau_R] - \log T^*. \qquad (11)$$

The production ratio ($P_R^{SNe} + (\Lambda_W / \Lambda_{SNe}) P_R^W$) / $P_S$ in Equation (11) is unknown where winds may be significant (i.e. where $P_R^W \neq 0$) as it reflects a number of factors (§4.3). In the absence of *a priori* information, we treat this production ratio as a fit parameter and invoke the ansatz that $\Lambda_W/\Lambda_{SNe}$ is sufficiently large that $^{26}Al/^{27}Al$ normalized by the production ratio lies on or near the array defined by $^{53}Mn$, $^{182}Hf$, $^{129}I$, $^{246}Sm$, $^{244}Pu$, $^{235}U$ and $^{238}U$ in Figure 1. Adopting the ansatz assumes also that production is sufficiently long-lived that $^{26}Al$ is continuously replenished and $\tau_{MC}$ therefore has meaning for short-lived nuclides as well as for the longer-lived nuclides. The validity of this assumption is evaluated in §4.4. A value of 26 for $(P^{SNe}_{^{26}Al} + (\Lambda_W/\Lambda_{SNe})P^W_{^{26}Al})/P_{^{27}Al}$ produces a value for $(N_{^{26}Al}/N_{^{27}Al})/((P^{SNe}_{^{26}Al} + (\Lambda_W/\Lambda_{SNe})P^W_{^{26}Al})/P_{^{27}Al})$ of $2 \times 10^{-6}$, in line with the aforementioned trend, as shown in Figure 3. Because with all else equal $P^W_{^{26}Al} \sim P^{SNe}_{^{26}Al}$ (Woosley and Heger 2007), the adopted production ratio implies $\Lambda_W/\Lambda_{SNe} \sim (26.0-0.017)/0.017 = 1530$ (Tables 1,2).

Wind production ratios for the other short-lived radionuclides are calculated from the expression

$$\frac{(\Lambda_W/\Lambda_{SNe})P^W_R}{P_S} = \frac{(\Lambda_W/\Lambda_{SNe})P^W_{^{26}Al}}{P_{^{27}Al}} \frac{P_{^{27}Al}}{P_S} \frac{P^W_R}{P^W_{^{26}Al}} \qquad (12)$$

where values for $P^W_R/P^W_{^{26}Al}$ are those for a 60 $M_\odot$ progenitor O star of solar metallicity (Arnould et al. 2006). Selection of production ratios for a single representative stellar mass is necessitated by the lack of IMF-integrated wind yields but is likely to be adequate given the uncertainties of 2x or more in production ratios in general and the limited range in production among larger O stars (e.g., figure 3 of Gounelle and Meynet, 2011). Stable isotope production ratios, $P_{^{27}Al}/P_S$, in Equation (12) are derived from the IMF-integrated, solar-normalized values of Woosley and Heger (2007) and the solar abundances (Lodders 2003). Production ratios from Equation (12) are listed in Table 2.

For those short-lived nuclides produced by WR winds, including $^{36}$Cl, $^{41}$Ca, $^{60}$Fe, and $^{107}$Pd, new values for $(N_R/N_S)/((P_R^{SNe}+(\Lambda_W/\Lambda_{SNe})P_R^W)/P_S)$, adjusted for the local wind contributions, are shown in Figure 3 plotted against $\tau_R$ (note that for radionuclides not produced by winds, $\Lambda P_R^W = 0$). Incorporation of pre-WR MS and WR winds results in all radionuclides aligning on a single trend consistent with $\tau_{MC} = 200$ +/− 100 Myr within errors (Figure 3). Most importantly, $^{36}$Cl, $^{41}$Ca, and $^{26}$Al, lying well above the $\tau_{MC} = 200$ curve in Figure 1, shift downward in Figure 3 by orders of magnitude while $^{60}$Fe, lying just above the curve in Figure 1, shifts just enough to settle onto the curve in Figure 3. Lastly, $^{107}$Pd, already on the curve in Figure 1, does not move when winds are included, thus preserving the alignment of all of the nuclides within error of a single $\tau_{MC}$ curve. The displacements in the radionuclide plotting positions in Figure 3 relative to Figure 1 are the result of additions in production spanning five orders of magnitude and serve to verify the ansatz that a single residence time ($\tau_{MC}$) applies to the solar system parental cloud. They suggest that winds played the dominant role in producing $^{36}$Cl, $^{41}$Ca, $^{26}$Al and made a subordinate contribution to $^{60}$Fe and a negligible contribution to $^{107}$Pd. Linear regression of the short-lived nuclides in Figure 3 for $\tau_R$ ranging from $^{41}$Ca to $^{244}$Pu gives a slope of 1.80 +/− 0.07 (1$\sigma$) (least squares with errors in abscissa and ordinate) and an intercept of −5.90 at log($\tau_R$) = 0. The best fit is within the +/− 100 Myr envelope in Figure 3.

## 4. DISCUSSION

*4.1 Consistency among radionuclide abundances*

Alignment of all radionuclides on a single curve for $\tau_{MC}$ of 200 +/−100 Myr in Figure 3 suggests that the solar abundances of short and long-lived nuclides are the result of processes described by Equation (11). The alignment is unlikely to be coincidental in that it involves products of stable isotope yields dominated by SNe (i.e., $P_{^{27}Al}^{SNe}/P_S^{SNe}$) and radioisotope yields dominated by winds of highly variable magnitude. If not coincidence, it follows that these nuclides were inherited from a typical molecular cloud 4.6 Gyr before present. While evidence for extinct short-lived radionuclides in the solar system has historically been taken as evidence for their anomalous concentrations, Figure 3 suggests instead that there are no resolvable excesses relative to expectations for a molecular cloud. For example, rather than being in excess, the solar relative abundance of $^{41}$Ca is more than 3 orders of magnitude *lower* than expected for a single-phase ISM (Figure 3). The relatively low concentrations of all of the short-lived radionuclides relative to expectations for a single-phase ISM are best explained as the result of slow exchange of mass between clouds and inter-cloud space characterized by a large value of $\tau_{MC}$ relative to decay mean lives. For the longer-lived radionuclides, this value for $\tau_{MC}$ affords free, unsupported decay. In the case of the shorter-lived radionuclides (SLRs), the value for $\tau_{MC}$ suppresses the abundances of SLRs by ensuring that significant radioactive decay occurs within MCs even as the masses of the nuclides are replenished (§4.3).

The alignment of radionuclides on a single array in Figure 3 suggests specific values for $\tau_{MC}$ and $\Lambda_W/\Lambda_{SNe}$. Both require independent assessments of plausibility.

*4.2 Value for $\tau_{MC}$*

The characteristic timescale $\tau_{MC}$ reflects the time to exit molecular clouds either by dissolution of the cloud entirely or conversion of a small fraction to stars and planets. The value for $\tau_{MC}$ in Figure 3 is an order of magnitude longer than the lifetimes of *individual* molecular cloud structures (e.g., Lada and Lada 2003; Murray et al. 2010). However, Elmegreen (2007) draws important distinctions between the lifetimes of cloud cores that produce star clusters, ~ 3 to 10 Myr, giant molecular clouds (GMCs), ~ 10 to 20 Myr, and super-cloud and cloud complexes, with lifetimes of > 50 Myr. These varying timescales are the result of the fact that more than 90% of cloud mass resides outside the densest regions where star formation is rife in surrounding envelopes. In the life cycle of giant molecular clouds, massive stars disrupt individual cloud masses in star-forming regions (Leisawitz et al. 1989; Elmegreen 2007), often before supernova formation (Murray et al. 2010; Pagani et al. 2012). However, this disruption is local, and the clouds are not so much completely destroyed as they are "shredded" (Elmegreen 2007). The inefficiency of star formation and the persistence of cloud fragments means that molecules and dust can persist in the molecular cloud phase for orders of magnitude greater timescales than the lifetime of a single cloud structure (Elmegreen 2007). The longevity of molecular cloud material is evidenced in spiral galaxy M51 where vestiges of GMCs are seen to traverse inter-arm space, requiring lifetimes of $10^8$ years (Koda et al. 2009). In a statistical sense, the principal means for exiting molecular clouds is therefore related to the rate at which stars form from clouds.

The rate of conversion of dense ISM to stars can be calculated from the molecular cloud mass, $M_{MC}$, of $8.4 \times 10^8$ $M_\odot$ and star formation rates, $\psi$, of 0.9 to 3 yr$^{-1}$ in the present-day Galaxy (Draine 2011). The mean timescale for converting clouds to stars is then $M_{MC}/\psi =$ 280 to 840 Myr with a median of 420 Myr. The median is within a factor of 2 of the $\tau_{MC}$ in Figure 3, suggesting that the solar radionuclide abundances are a reflection of the rate of star formation from molecular clouds ca. 4.6 Gyr before present. It is noteworthy that the characteristic timescale for dust destruction in the ISM is also estimated to be ~400 Myr to 600 Myr Tielens (2005).

*4.3 Value for $\Lambda_W/\Lambda_{SNe}$*

Values for $\Lambda_W/\Lambda_{SNe} \gg 1$ are implicit in the conclusion that $^{26}$Al appears to be mainly concentrated in clouds (Jura et al. 2013). In addition, the consistency of $^{26}$Al concentrations (i.e., $^{26}$Al/$^{27}$Al) in star-forming regions and resemblances to early solar values implies recurring local production near clouds (Jura et al. 2013). This recurring production is simulated in §4.4. Figure 3 suggests that not only is $^{26}$Al produced with $\Lambda_W/\Lambda_{SNe} \gg 1$, but that $^{36}$Cl and $^{41}$Ca are as well. In contrast, $^{53}$Mn, $^{60}$Fe, $^{107}$Pd, $^{182}$Hf, $^{129}$I, $^{244}$Pu, $^{146}$Sm, and the longer-lived radionuclides, are evidently produced mainly by supernovae. Localized production of the shortest-lived radionuclides (those with $\tau_R \leq 1$Myr) should result in a difference between average Galactic ratios of shorter and longer-lived nuclides and the ratios observed in materials produced in active star-forming regions (e.g., the solar system).

Therefore, one measure of the focusing of production in star-forming regions (SFRs) by winds might be the atomic steady-state $N_{60_{Fe}}/N_{26_{Al}}$ of 0.55+/− 0.22 for the Galaxy as a whole implied by γ emission (Wang et al. 2007; Tang and Dauphas 2012), compared with the $N_{60_{Fe}}/N_{26_{Al}}$ in SFRs represented by the solar value of 0.002. If SFRs sample $^{60}$Fe/H of the diffuse ISM with fidelity, an assumption that is far from assured, then $(^{60}Fe/^{26}Al)_{Galaxy}/(^{60}Fe/^{26}Al)_\odot = 300$ +/− 110, implying that $\Lambda_W/\Lambda_{SNe} \sim 300$. The uncertainties in this analysis are difficult to quantify but include uncertainties in production ratios that are typically known only to factors of 2 to 3 and the uncertainty surrounding the assumption that $^{60}$Fe is uniformly distributed throughout all phases of the ISM. Using $\Lambda_W/\Lambda_{SNe}$ = 300 leaves $^{41}$Ca still within error of the 200 Myr $\tau_{MC}$ trend in Figure 3 and moves $^{36}$Cl, $^{26}$Al, and $^{60}$Fe more than a factor of 2 above the trend.

A number of factors may contribute to $\Lambda_W/\Lambda_{SNe} \gg 1$. The life expectancies of massive B and O stars, $\tau_*$, can be calculated as $\tau_* = 10^{-6} (1.13 \times 10^{10} M_*^{-3} + 0.6 \times 10^8 M_*^{-0.75} + 1.2 \times 10^6)$ with stellar mass $M_*$ in solar mass units (Schaller et al. 1992; Prantzos 2007). These values for $\tau_*$ with a typical IMF (e.g. the Salpeter IMF) show that 75% of supernovae progenitor stars will outlive their parental clouds if the latter persist for $\leq 10$ Myr (75% of supernova progenitor stars are B stars with $8 M_\odot < M_* < 18 M_\odot$). With a typical velocity dispersion of 10 km/s (Elmegreen 2007), the larger B stars will travel ~100 pc (of the order of the Galactic disk scale height) from their birth environment and their parental cloud will have largely dissipated by that time. All else equal, a 30 $M_\odot$ O star (~ minimum for a WR progenitor) that evolves to a WR star will move on average ~60 pc during its lifetime of 6 Myr (an ~ maximum for WR stars) and the parental cloud is likely to be intact. The small fraction of supernovae exploding during the lifetimes of parental clouds (~0.25) and the reduction in flux caused by increasing distances leads to a reduction in the efficiency of SNe ejecta capture. However, any geometric and/or temporal biases towards enrichments by winds relative to supernovae will be diluted by subsequent explosion of wind-producing stars as collapse SNe. A more fundamental means for accruing a high fraction of wind products relative to SNe ejecta in star-forming regions is evidently required.

Wolf-Rayet stars that collapse more or less passively to black holes offer a natural explanation for high $\Lambda_W/\Lambda S_{Ne}$. There is evidence that many stars with $M_* > \sim 25$ to $35 M_\odot$ and with metallicities of solar or less do not produce bright and energetic SNe but rather collapse by fallback to form black holes directly with either no associated supernova (prompt black hole formation) or perhaps just a dim, failed supernova (delayed black hole formation) (Smartt 2009). Where sufficiently massive, these stars may be important sources of gamma-ray bursts ("collapsars") (Woosley 1993; Fryer 1999; Woosley et al. 1999; Fryer et al. 2007; Langer 2012). Direct formation of black holes is likely to be more common for single massive WR stars. The Type Ib/c supernovae traditionally attributed to WR progenitors (due to the paucity of hydrogen) are probably more often produced by lower-mass helium stars robbed of hydrogen by a binary companion and with initial masses ranging below the WR minimum threshold (Smartt 2009; Langer 2012; Eldridge et al. 2013). The overall conclusion is that many (up to 95%, Smartt 2009) WR stars may emit substantial winds without subsequent ejection of SNe debris. An example is thought to be Cygnus X-1. Cyg X-1 is a black hole formed from a WR progenitor with an initial mass > 40 $M_\odot$ (Mirabel and Rodrigues 2003). The formation of this black hole evidently occurred with an undersized explosion or with no explosion at all (Mirabel and Rodrigues 2003). Counter examples also exist (Groh et al. 2013) and the frequency of supernovae produced by WR stars remains uncertain and a subject of active investigation.

*4.4 Model for* $\Lambda_{SNe} P_{^{26}Al}^{SNe} + \Lambda_W P_{^{26}Al}^W$

As an illustration of the potential for protracted production of short-lived nuclides, the production of $^{26}$Al adjacent molecular clouds in star-forming regions was simulated numerically (Appendix). $^{26}$Al from winds and/or supernovae produced within the 10 Myr lifespan of their parental clouds were summed over multiple generations of stellar clusters. Decay of $^{26}$Al was tracked simultaneously (Appendix). These episodes of star formation were placed in series to represent hundreds of millions of years of evolution comprising the long-term production of $^{26}$Al (i.e., $\Lambda_{SNe} P^{SNe}_{^{26}Al} + \Lambda_W P^W_{^{26}Al}$) resulting from successive episodes of star formation. Yields for $^{26}$Al were obtained by interpolation of published models as a function of progenitor mass (Appendix) for both WR winds and collapse supernovae. Two scenarios were considered. In one, all WR stars explode as collapse SNe. In the other, WR stars (minimum mass of ~ 25 $M_\odot$) with progenitor masses exceeding 28 $M_\odot$ do not explode (as if they collapse passively to black holes instead). These two scenarios represent end-members for comparison. Results are shown in figures 4 and 5.

In Figure 4 the time evolution of the mass of $^{26}$Al available for molecular clouds experiencing multiple episodes of star formation are shown for the case where WR stars end as supernovae and the case where WR stars > 28 $M_\odot$ do not form supernovae. In both cases the abundances of $^{26}$Al are irregular but reach a quasi-steady state near the typical yield for WR winds and SNe (~ $10^{-4}$ to $10^{-5}$ solar masses of $^{26}$Al). This justifies the use of long-term, persistent production in Equation (11) for radionuclides with mean lives orders of magnitude shorter than the life span of cloud material (in this case 1 Myr vs. 400 Myr, respectively); these nuclides are constantly replenished by the star formation process. This phenomenon is analogous to the persistence of short-lived radionuclides in secular equilibrium in decay chains of long-lived isotopes (e.g., persistence of $^{266}$Ra, $\tau_{^{266}Ra}$ = 2307 yrs, in the $^{238}$U decay series with $\tau_{^{238}U}$ = 6.4x10$^9$ yrs).

While the two plots in Figure 4 resemble one another, the ratio of $^{26}$Al produced by WR stars vs. SNe is markedly different in the two cases, as shown in Figure 5. Since masses of $^{26}$Al produced by winds and by supernovae are similar for a given progenitor mass, the ordinates in Figure 5 are effectively $\Lambda_W/\Lambda_{SNe}$. In the case where all stars die as supernovae, supernova yields tend to dominate the inventory of short-lived nuclides (e.g., $^{26}$Al) (Figure 5, left panel). However, where massive O stars that evolve inevitably to WR stars end without supernova explosions, more often than not winds dominate the yields (Figure 5, right panel). In these cases $\Lambda_W/\Lambda S_{Ne}$ can be as high as $10^3$ for significant periods of time with maximum values of ~ $10^7$ to $10^8$. Figures 4 and 5 suggest that there are plausibly common circumstances where winds dominate over supernova ejecta in star-forming regions. The effect of the fate of O stars on $\Lambda_W/\Lambda S_{Ne}$ is depicted schematically in Figure 6.

*4.5 Mixing and mean life*

The production models in Figures 4 and 5 are for $^{26}$Al. The mean lives of $^{41}$Ca and $^{36}$Cl are even shorter than that for $^{26}$Al. We would expect, therefore, that these nuclides would deviate more strongly from the best-fit curve in Figure 3 due to the finite timescales of mixing in clouds relative to the rate of radioactive decay and free decay between recurrent injections. In lieu of a detailed numerical simulation of star-forming regions that includes WR winds, shredding and reassembly of molecular clouds, and cloud mixing (beyond the scope of this work), we can crudely account for the effects of incomplete mixing in Figure 3 by invoking the phenomenological equation for concentration dispersion (σ) as a function of characteristic mixing time $\tau_{mixing}$ and decay mean life (Junge 1974; Jobson et al. 1999):

$$\sigma_{\log x} = 0.43 \left( \frac{\tau_{mixing}}{\tau_R} \right)^{1/2} \qquad (13)$$

where in this case *x* is the ordinate in Figure 3 and 0.43 is log(*x*)/ln(*x*). Mixing between and within clouds takes place over parsec scales ($3 \times 10^{16}$ m) at ~ sound speed $C_s = \sqrt{kT/\mu}$. For cloud temperatures *T* of 10 k and reduced mass $\mu = 2.7 \times 1.67 \times 10^{-21}$ kg, $C_s = 175$ m s$^{-1}$. We can calculate a mixing timescale from $3 \times 10^{16}$ m/175 m s$^{-1}$ = $1.7 \times 10^{14}$ s or 5 Myr. This figure compares favorably with cloud turbulent velocities yielding $\tau_{mixing} \sim 2$ Myr (Xie et al. 1995). Curves based on this mixing timescale and Equation (13) are shown in Figure 3 bracketing the best-fit molecular cloud residence time, depicting the expected dispersion in relative short-lived radionuclide abundances as a function of $\tau_R$.

The $^{41}$Ca and $^{36}$Cl data adhere more tightly to the best fit for the specified value of $\Lambda_W/\Lambda S_{Ne}$ than the delays due to mixing might imply. If the ansatz invoked above is correct, this implies that production from winds is more continuous than presently accounted for and/or inter-cloud and intra-cloud mixing is more vigorous than that described here.

*4.6 Testing predictions with astronomical observations*

High $\Lambda_W/\Lambda S_{Ne}$ leads to the prediction that the positions of WR stars should be well correlated with the positions of molecular clouds in star-forming regions while the locations of supernovae should be less well correlated. This prediction arises for two reasons. Firstly, if massive O stars generally go through the WR phase but do not inevitably explode vigorously as supernovae, there should be a near perfect correlation between occurrences of WR stars and molecular clouds and less of a correlation between SNe and clouds. Secondly, SNe that form from less massive B stars can explode further removed from star-forming regions than their O-star siblings.

In principle, this prediction is testable by comparing the spatial distributions of WR stars, SNe remnants, and molecular clouds. In practice, this comparison is difficult within the Galaxy because while galactic longitude and latitude are well constrained, confusion in the third dimension arises from uncertainties in distances that are significant when viewing the Galactic disk edge on. A better test should come from viewing galaxies face on.

The nearly face-on Local Group spiral galaxy M33 (the Triangulum Galaxy) provides an opportunity to compare distributions of WR stars (n=206, Neugent and Massey 2011), SNe remnants (n=137, Long et al. 2010), and molecular clouds (n=149, Rosolowsky et al. 2007) because the uncertainties in distances are no more than the ~ scale height of the galactic disk. A two-dimensional Kolmogorov-Smirnov (K-S) test was used to assess the probability that the positions of WR stars, molecular clouds, and supernova remnants in M33 represent the same 2-D distribution. The test follows procedures outlined by Press et al. (Press et al. 2007, Chapter 14). For these tests the $D$ statistics (a measure of the maximum disparity in fractions of data present around each datum from each of two data sets being compared) for pairs of data sets were compared with Monte Carlo simulations of the data. The latter establish the distribution of $D$ for the null hypothesis where the distributions are the same. A Gaussian (as opposed to a uniform) distribution was used to produce the fictive random data to account for the tendency for all objects to be concentrated towards the center of the galaxy. One thousand synthetic simulations of the data sets were obtained by random draws, each with the same number of data points and same mean and standard deviations in Galactic longitude and latitude as the actual data. Probabilities for correlation between two groups were obtained by counting the fraction of synthetic $D$ statistics for random data that exceed the $D$ statistic for the actual data.

Results of the K-S tests show that there is a 72% probability that the WR stars in M33 have the same 2-D distribution as that defined by molecular clouds in M33. The probability that SNe remnants have the same distribution as clouds drops to 56%. The probability that WR stars and SNe remnants share the same 2-D distribution is 27%. While not dispositive, these first results are consistent with WR stars evolving in the vicinity of molecular clouds in star-forming regions and SNe exploding further away from clouds.

5. CONCLUSIONS

Inclusion of winds from massive O stars can lead to alignment of short and long-lived radio nuclides along a well-defined trend in $\log[(N_R/N_S)/(P_R/P_S)]$ vs. $\log(\tau_R)$ space when supernova production ratios are augmented by production ratios that include WR winds. The trend indicates that the solar abundances of radionuclides could have simply been inherited from the solar parental molecular cloud. This cloud had abundances of radionuclides consistent with the average residence time of $10^8$ years indicated by the present-day rate of converting molecular clouds into stars in the Milky Way. The apparent excesses in $^{26}$Al, $^{36}$Cl, $^{41}$Ca, and to a lesser extent $^{60}$Fe, relative to other short- and long-lived nuclides, may well be artifacts of omitting the effects of massive-star winds preferentially deposited into dense phases of the ISM in star-forming regions. This analysis suggests that the solar concentrations of these extinct isotopes in the early solar system do not *a priori* signify injection of specific proximal stellar sources, but instead reflect average concentrations in average star-forming clouds.

The alignment of the short-lived nuclides upon adjustment for winds is difficult, if not impossible, to explain by chance given that the wind production terms for these nuclides vary by 5 orders of magnitude. Still, this analysis relies on uncertain wind production and capture by clouds relative to other sources of the elements in star-forming regions, and verification of this hypothesis will require a better understanding of the heterogeneous production and capture of nuclides in different phases of the interstellar medium. Conversely, the solar abundances of shortest-lived nuclides may be telling us that the most massive stars in star-forming regions can collapse without supernova ejection.

ACKNOWLEDGMENTS

The author is indebted to Mike Jura for on-going discussions on the topics of heterogeneous distribution of radionuclides and star formation. Matthieu Gounelle, Francis Nimmo, Steven Desch, Brad Meyer, and Stein Jacobsen all freely shared their time and insights with the author on the topic of short-lived radionuclides. Gounelle and Larry Nittler provided helpful critical reviews of a previous version of the manuscript. I thank Michael Shara for suggesting the statistical test in M33. Funding was provided by the NASA Cosmochemistry program.

APPENDIX: PRODUCTION MODEL

An IDL+Fortran code was written to simulate the production of $^{26}$Al in the vicinity of molecular cloud material for prolonged periods of time (~400 Myrs). The goal is to track the availability of $^{26}$Al to cloud material that survives multiple episodes of star formation, as described by Elmegreen (2007). Fluxes between the diffuse interstellar medium and molecular clouds are not modeled here; this simulation represents only the long-term production term in Equation (11).

Fictive star clusters were generated using the mass generation function of Kroupa et al. (1993) modified by Brasser et al. (2006). With this function, initial masses for each star $j$ are drawn at random according to the expression

$$M_j / M_\odot = 0.01 + (0.19 x^{1.55} + 0.05 x^{0.6}) / (1-x)^{0.58} \qquad (A.1)$$

where $x$ is a uniformly distributed random number between 0 and 1. For the results in this paper we used 5000 stars per cluster (i.e., 5000 random draws for $x$), representing the high end of cluster populations where WR stars (i.e., O stars) are most likely to be found. In the simulation, each cluster is considered to be proximal to a parental cloud for 10 Myr. This cloud material is then exposed to successive generations of star formation; the simulation consists of tens of clusters produced in series. For example, for 30 clusters, the total duration of the simulation is 300 Myr (10x30). In order to allow for a finite interval of star formation (i.e., all stars in a cluster are not born at exactly the same time), the birth date of each star in a cluster was altered at random over a total time interval of 1 Myr. The exact value of this "blurring" of birth dates turns out to have little effect on the models.

The minimum mass for WR activity is ~ 25 $M_\odot$, meaning that all WR stars eject winds in the vicinity of cloud material in this model. For 5000 stars per cluster in these simulations, the average number of WR stars per cluster is 1.6 +/− 1.2 (1$\sigma$, based on sampling 500 clusters), all of which evolve and die within 10 Myr of their birth. The minimum initial mass for a star to deposit supernova ejecta proximal to clouds in the simulations is 18 $M_\odot$ (based on stellar life times, see text), meaning that only ~24% of supernovae provide ejecta in these simulations. The average number of supernovae is 11.9 +/− 3.2 while the average number of supernovae exploding within 10 Myr of birth is 2.9 +/− 1.7.

The mass of $^{26}$Al ejected by WR winds and supernovae proximal to cloud material is tallied as a function of time. Integrated masses ejected by WR winds were obtained in approximate form from Gounelle and Meynet (Gounelle and Meynet 2012). The total yields as a function of stellar progenitor masses used for interpolation for all masses are

| Progenitor Mass ($M_\odot$) | $^{26}$Al ($M_\odot$) |
|---|---|
| 20 | 2.0x10$^{-10}$ |
| 25 | 2.0x10$^{-6}$ |
| 30 | 2.0x10$^{-5}$ |
| 60 | 7.0x10$^{-5}$ |
| 80 | 1.0x10$^{-4}$ |
| 120 | 2.0x10$^{-4}$ |

where the maximum stellar mass considered was 120 $M_\odot$. Yields for Supernovae were obtained from interpolation of values given by Chieffi and Limongi (2013). Precision in yields is not important to the conclusions to be drawn from this simulation.

For every time step (~0.3 Myr time intervals) each parcel of $^{26}$Al delivered by a star that experiences winds or ends its life as a supernova within 10 Myrs of the initiation of the cluster was added to the total inventory of available $^{26}$Al for the duration of the model (usually ~ 400 Myr in this work, although in practice these parcels are of little consequence after > 10 half-lives). For supernova ejecta a single pulse of $^{26}$Al with mass $M^{SNe^o}_{^{26}Al,j}$ is added to the total inventory of $^{26}$Al at the moment the star explodes (if the star does so within 10 Myr of the birth of its cluster). The radioactive decay of this parcel of $^{26}$Al from supernova $j$ is followed through time $t$ using

$$M_{^{26}Al,j} = M^{SNe^o}_{^{26}Al,j} \exp(-t/\tau_{^{26}Al}). \tag{A.2}$$

Products of winds evolve in a more complicated fashion because they are delivered over extended periods of time, with time constant $\tau_W$, simultaneous with radioactive decay with mean life $\tau_{^{26}Al}$. Their evolution is therefore governed by the equation

$$\frac{dM_{^{26}Al,j}}{dt} = \frac{M^W_{^{26}Al}}{\tau_W} - \frac{M_{^{26}Al,j}}{\tau_{^{26}Al}} \tag{A.3}$$

where $M_{^{26}Al,j}$ is the time-dependent mass of $^{26}$Al evolving from stellar source $j$ as modified by radioactive decay and $M^W_{^{26}Al,j}$ is the evolving mass of $^{26}$Al derived from the stellar source $j$ at time $t$. The solution for each WR source $j$ is

$$M_{^{26}Al,j}(t) = \frac{\tau_{^{26}Al}}{\tau_W - \tau_{^{26}Al}} M^{W^o}_{^{26}Al}(\exp(-t/\tau_W) - \exp(-t/\tau_{^{26}Al})) \tag{A.4}$$

where $M^{W^o}_{^{26}Al,j}$ is the total integrated mass of $^{26}$Al delivered by winds from source $j$. Inspection of the time evolution of wind products given by Gounelle and Meynet (2012) suggests a value for $\tau_W$ of ~ 8 Myr. Many O stars end their lives prior to this time interval, so the time evolution in Equation (A.3) is cut short with the death of the WR star and is replaced by simple decay of the remaining $^{26}$Al using the form of Equation (A.2). Typical results for WR winds only and for winds together with supernovae are shown in Figure A1.

**Table 1**. Data used to construct Figure 1.

| Species | $N_R/N_S$ | | $P_R^{SNe}/P_S$ † | | Mean Life (Myrs)¶ | |
|---|---|---|---|---|---|---|
| $^{26}$Al/$^{27}$Al | 5.2x10$^{-5}$ | (1) | 1.7x10$^{-2}$ | (12) | 1.05 | (14) |
| $^{36}$Cl/$^{35}$Cl | 5.0x10$^{-6}$ | (2) | 2.5x10$^{-2}$ | (12) | 0.43 | (12) |
| $^{41}$Ca/$^{40}$Ca | 4.2x10$^{-9}$ | (3) | 2.3x10$^{-3}$ | (12) | 0.15 | (15) |
| $^{53}$Mn/$^{55}$Mn | 9.1x10$^{-6}$ | (4) | 0.83 | (12) | 5.34 | (16) |
| $^{60}$Fe/$^{56}$Fe | 1.15x10$^{-8}$ | (5) | 1.23x10$^{-4}$ | (12) | 3.78 | (17) |
| $^{107}$Pd/$^{108}$Pd | 5.9x10$^{-5}$ | (6) | 0.84 | (12) | 9.38 | (18) |
| $^{129}$I/$^{127}$I | 1.0x10$^{-4}$ | (7) | 1.4 | (12) | 23.0 | (19), (13) |
| $^{146}$Sm/$^{144}$Sm | 8.0x10$^{-3}$ | (8) | 2.7 | (12) | 98.10 | (20) |
| $^{182}$Hf/$^{180}$Hf | 1.0x10$^{-4}$ | (9) | 0.33 | (12) | 12.84 | (21) |
| $^{244}$Pu/$^{232}$Th | 3.0x10$^{-3}$ | (10) | 1.14 | (12) | 115.0 | (22), (12) |
| $^{235}$U/$^{232}$Th | 0.133 | (11) | 1.11 | (13) | 1015. | (23) |
| $^{238}$U/$^{232}$Th | 0.415 | (11) | 0.84 | (13) | 6446. | (23) |

† Uncertainties in production ratios taken to be +2$x$ /-1/2$x$ where $x$ is the value for the production ratio. These values are IMF-integrated supernova-dominated production ratios.

¶ Uncertainties in mean lives taken to be +/- 10%.

Data sources: (1) Jacobsen et al. (2008); (2) Lin et al. (2005); (3) Liu et al. (2012); (4) Nyquist et al. (2009); (5) Tang and Dauphas (2012); (6) Schönbachler et al. (2008); (7) Brazzle et al. (1999) ; (8) Stewart et al. (1994); (9) Kleine et al. (2005); (10) Hudson et al. (1989); (11) Wasserburg et al. (1996); (12) Huss et al. (2009); (13) Jacobsen (2005); (14) Norris et al. (1983); (15) Paul et al. (1991); (16) Honda and Imamura (1971); (17) Rugel et al. (2009); (18) Flynn and Glendenin (1969); (19) Russell (1957); (20) Kinoshita et al. (2012); (21) Vockenhuber et al. (2004); (22) Bemis et al. (1969); (23) Steiger and Jäger (1977).

**Table 2**. Wind mass production terms and atomic production ratios used to construct Figure 3.

| Ratio | $AP_R^W$ † | $P_{27_{Al}}^W/P_S$ | $(\Lambda_W/\Lambda_{SNe})P_R^W/P_S$ †† | $(P_R^{SNe}+(\Lambda_W/\Lambda_{SNe})P_R^W)/P_S$ ¶ |
|---|---|---|---|---|
| $^{26}Al/^{27}Al$ | $7\times10^{-5}\ M_\odot$ (1) | 1.0 | 25.98 | 26.00 |
| $^{36}Cl/^{35}Cl$ | $7\times10^{-7}\ M_\odot$ (2) | 28.26 (3) | 5.303 | 5.328 |
| $^{41}Ca/^{40}Ca$ | $8\times10^{-7}\ M_\odot$ (2) | 2.21 (3) | 0.4158 | 0.4181 |
| $^{60}Fe/^{56}Fe$ | $3\times10^{-9}\ M_\odot$ (2) | 0.194 (3) | $9.361\times10^{-5}$ | $2.170\times10^{-4}$ |
| $^{107}Pd/^{108}Pd$ | $9\times10^{-10}\ M_\odot$ (2) | 0.800 (3) | $6.494\times10^{-5}$ | 0.8429 |

† $A$ = atomic mass ratio and converts $P_R^W$ in atomic fraction to solar mass, other columns are in atomic units.

Sources: (1) Gounelle and Meynet (2012); (2) Arnould et al. (2006); (3) Woosley and Heger (2007).

†† Calculated from Equation (12) with $\Lambda_W/\Lambda_{SNe}\ P_{26_{Al}}^W/P_S = 25.98$.

¶ Total production ratios (Equation (12)) using the SNe production ratios in Table 1 and column 4 of this table.

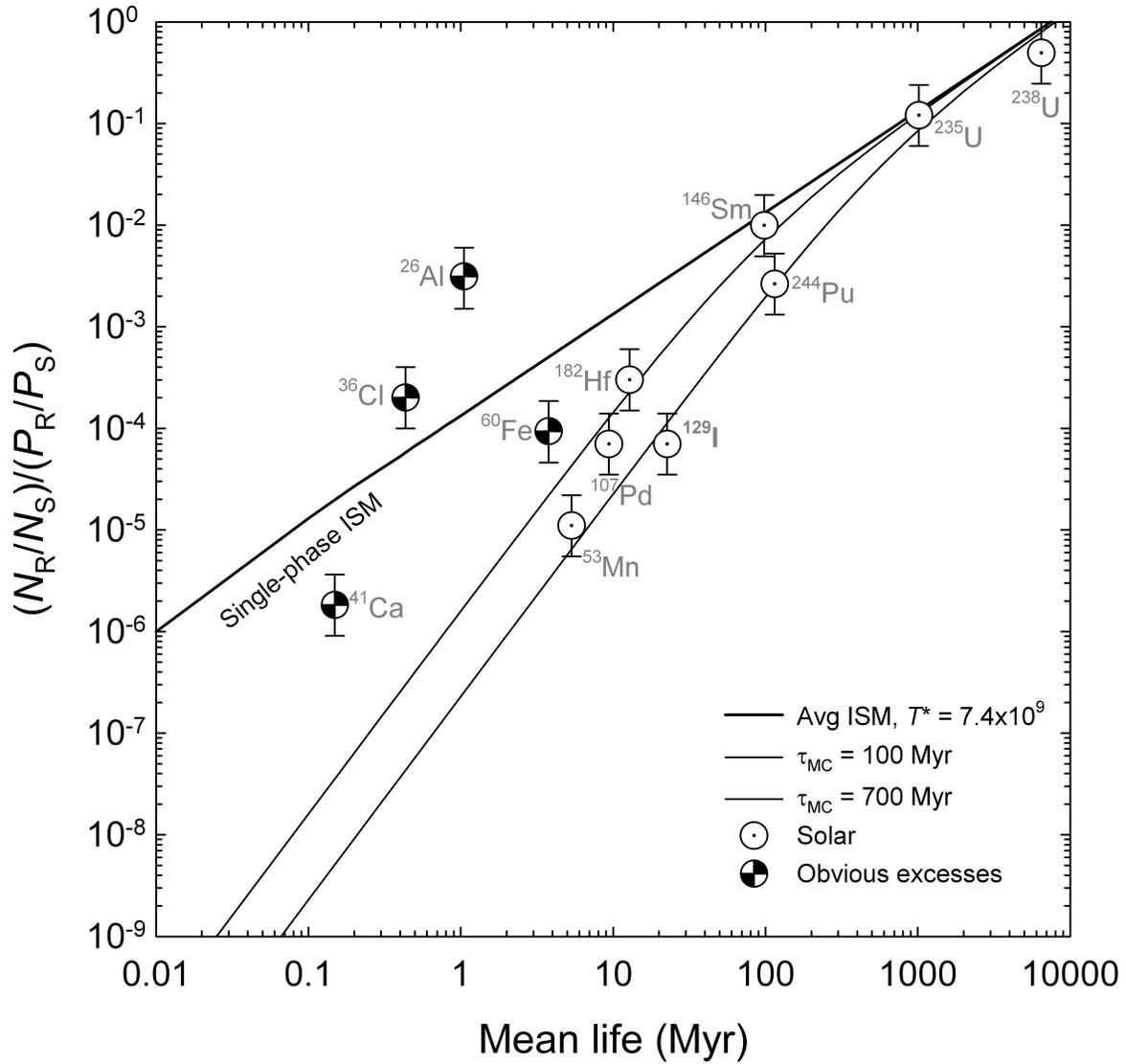

**Figure 1**. Plot of radioisotope/stable isotope ratios normalized to supernova-dominated production ratios vs. mean lives. Open solar symbols are radionuclides where winds from massive stars do not appear to be important. Black/white symbols refer to radionuclides that exhibit apparent excesses relative to predictions for a two-phase ISM. The curve labeled single-phase ISM represents expectations where there is no opportunity for decay without production (Equation 2). The two lower curves represent residence times in molecular clouds ($\tau_{MC}$) where they are isolated from supernova production (Equation 7). Only the radionuclides are labeled. The isotope ratios implied by the labels are shown in Table 1. Errors bars in the abscissa (mean life) are smaller than the symbols (see Table 2).

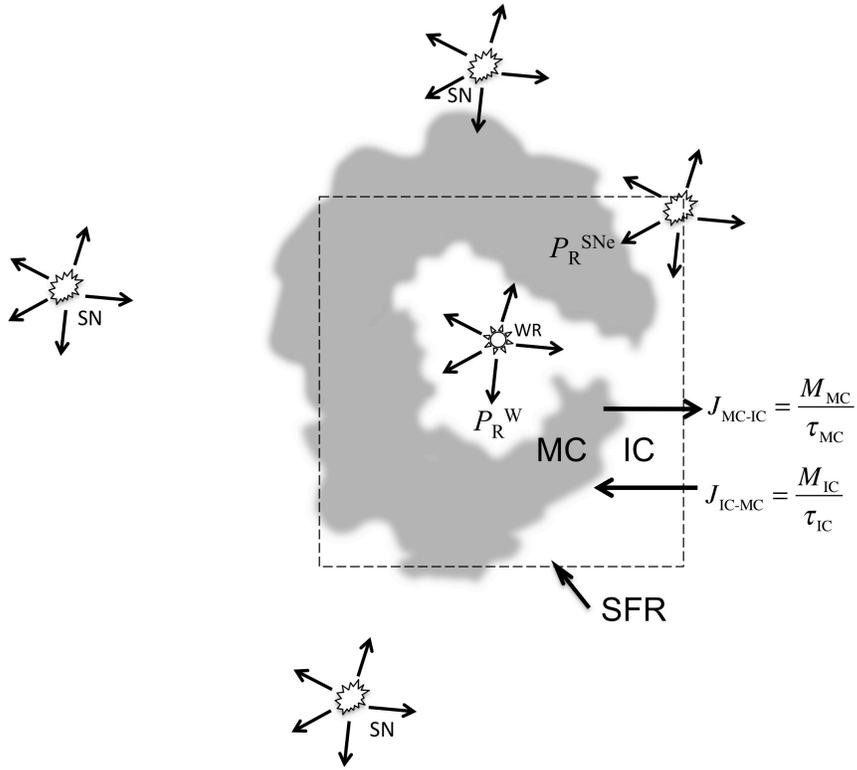

**Figure 2**. Schematic illustration of the factors included in Equations (3-6) and Equations (9) and (10). Production of short-lived radionuclides R from WR stellar winds ($P_R^W$) and supernovae ($P_R^{SNe}$) are indicated. The relationships between star-forming regions (SFR) composed of new stars and molecular clouds (MC), inter-cloud space (IC), and fluxes $J$ between them are shown in terms of the masses of the reservoirs $M_{MC}$ and $M_{IC}$ and the residence times in those reservoirs $\tau_{MC}$ and $\tau_{IC}$. Low fluxes correspond to large residence times.

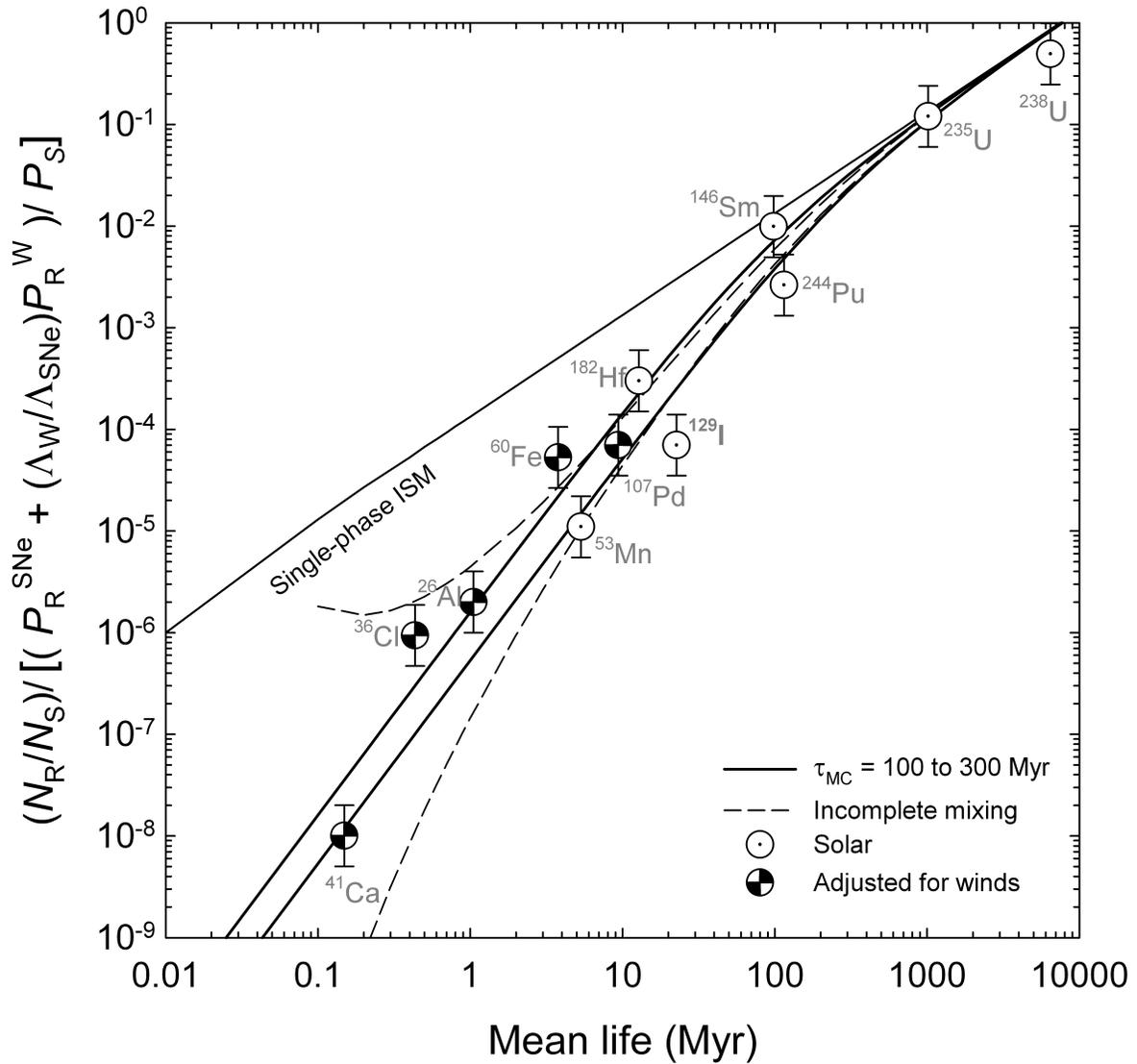

**Figure 3**. Same as Figure 1 but with production from WR and pre-WR winds added to the denominator of the ordinate values for the data. $^{26}$Al, $^{36}$Cl, $^{41}$Ca, $^{60}$Fe, and $^{107}$Pd are shifted as a result of the added wind production. The shift for $^{26}$Al is *assumed* by ansatz while the others are calculated based on the ratios of yields from Arnould et al. (2006). Incorporation of wind production removes the apparent excesses exhibited by $^{26}$Al, $^{36}$Cl, $^{41}$Ca, and $^{60}$Fe without resulting in a deficit for $^{107}$Pd. Nuclides shown with the open solar symbols are not produced in the winds. Curves bracketing a 200 +/− 100 Myr residence time ($\tau_{MC}$) are shown for reference and provide a reasonable fit to the adjusted data. Dashed line shows the dispersion expected from incomplete mixing in clouds (see text).

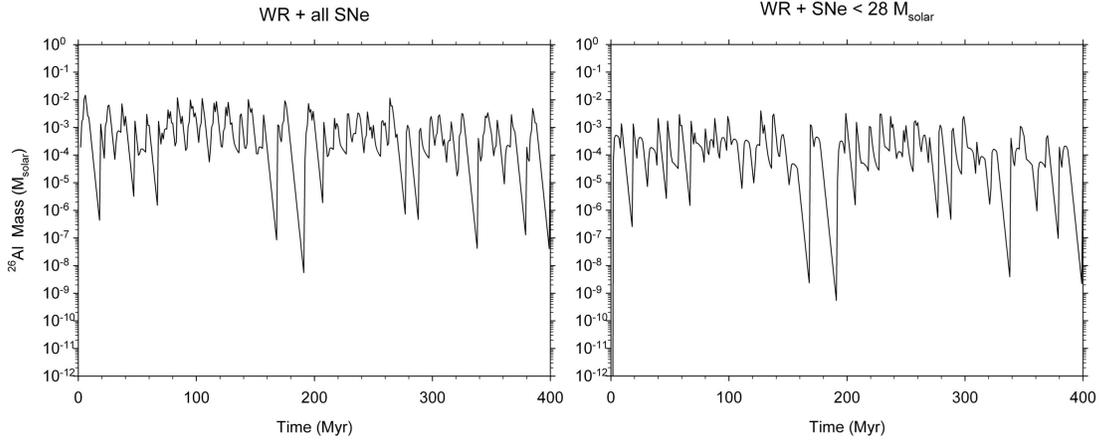

**Figure 4**. Results of simulations of $^{26}$Al production by multiple star-forming events. Each plots shows the mass of $^{26}$Al ejected, modified by continuous radioactive decay. *Left:* all WR stars evolve through the supernova stage. *Right*: only stars with progenitor masses < 28 solar masses evolve to form supernovae.

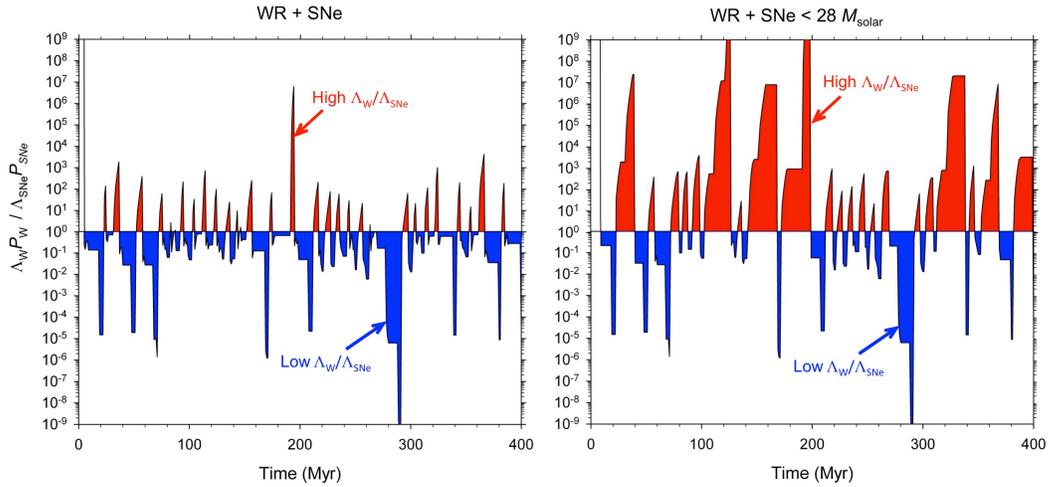

**Figure 5**. Ratio of $^{26}$Al production by WR stars to that by SNe in the simulations shown in Figure 4. Blue signifies dominance by SNe ($\Lambda_W/\Lambda_{SNe} < 1$) and red signifies dominance by WR winds ($\Lambda_W/\Lambda_{SNe} > 1$). Where supernova formation is limited to stars < 28 solar masses (*right*), winds can dominate more often than not.

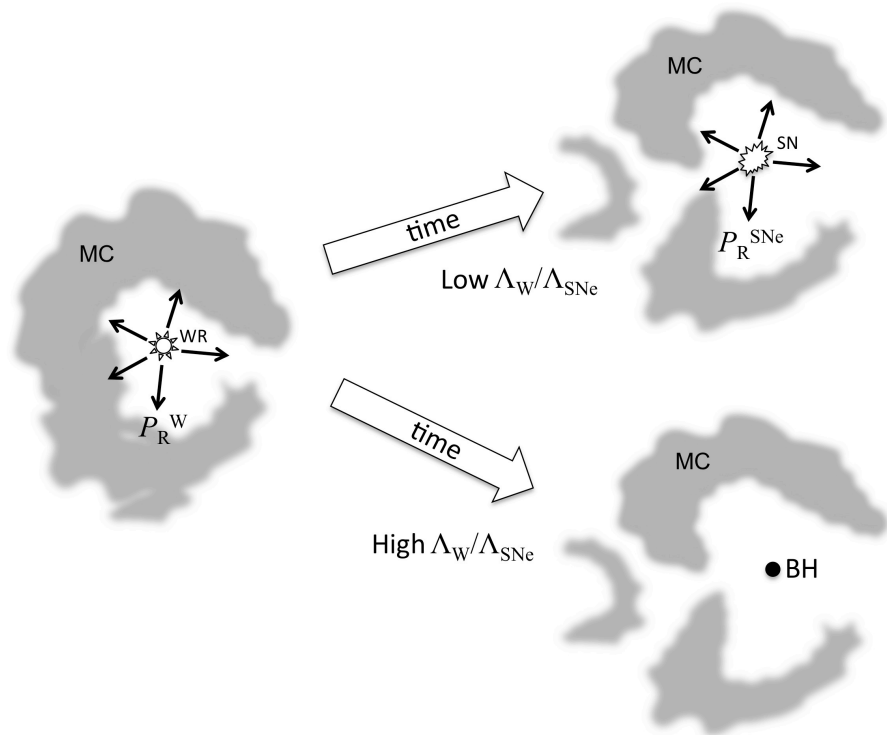

**Figure 6**. Schematic illustration of the possible evolution of WR stars to either supernovae (SN) or black holes (BH) without vigorous supernova explosion and the effect on $\Lambda_W/\Lambda_{SNe}$.

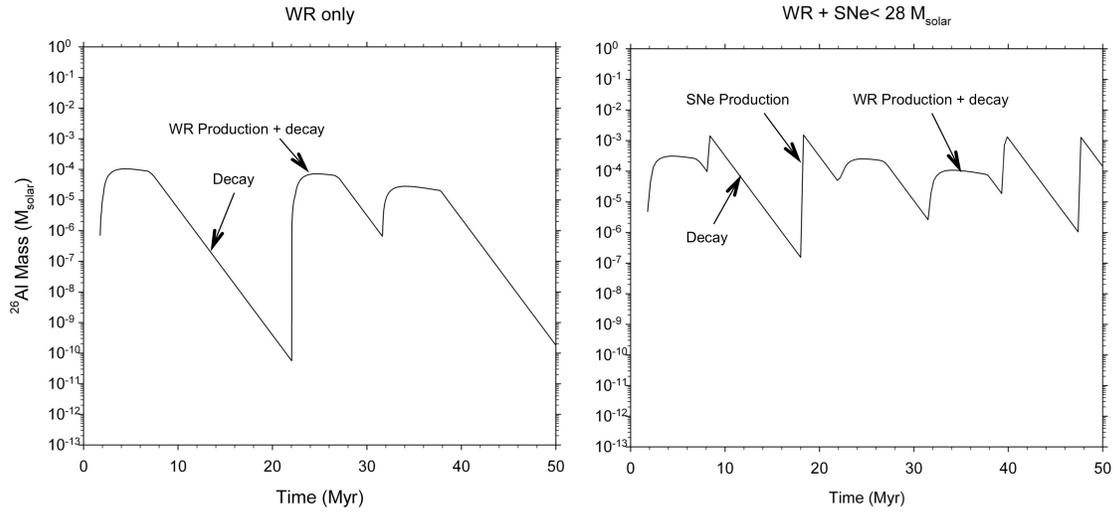

**Figure A1**. Examples of WR and WR+SNe < 28 $M_\odot$ $^{26}$Al production for a star-forming region. *Left*: WR production for three successive WR stars using Equations A.4 prior to the death of the star and A.2 after termination. *Right*: Same as at left with the addition of four supernovae where $^{26}$Al is added instantaneously followed by decay using Equation A.2.